\documentclass[sigconf]{webmedia}

\AtBeginDocument{%
  \providecommand\BibTeX{{%
    \normalfont B\kern-0.5em{\scshape i\kern-0.25em b}\kern-0.8em\TeX}}}

\setevent{wfa}
\proceedingsDetails[WFA’2025]{XXIV Workshop de Ferramentas e Aplica{\c c}{\~o}es (WFA 2025). Anais Estendidos do XXXI Simpósio Brasileiro de Sistemas Multimídia e Web}{2025}{Rio de Janeiro/RJ, Brasil}
\ISSN{2596-1683}

\usepackage[english,brazil]{babel}
\usepackage[T1]{fontenc}
\usepackage[utf8]{inputenc}

\usepackage{graphicx}
\usepackage{amsfonts}   
\usepackage{amstext}    
\usepackage{amsmath}    
\usepackage{mathtools}  
\usepackage{calc}       
\usepackage{setspace}     
\usepackage{verbatim}     
\usepackage{textcomp}     
\usepackage{soul}         
\usepackage{xfrac}        
\usepackage{icomma}  
\usepackage{paralist}  
\usepackage{relsize}  
\usepackage{enumitem}
\usepackage{afterpage}  
\usepackage{fancyvrb}  
\usepackage{color}
\usepackage{lscape}  
\usepackage{hhline} 
\usepackage{layout}  
\usepackage{fancyhdr}  
\usepackage{stfloats}  
\usepackage{longtable}  
\usepackage{multirow}   
\usepackage{hhline}     
\usepackage{array}      
\usepackage{booktabs}   
\usepackage{ctable}  
\usepackage{colortbl}  
\usepackage{subfigure}  
\usepackage{float}      
\usepackage{graphics,graphicx}  
\usepackage{epstopdf}  
\usepackage{psfrag}  
\usepackage{wallpaper}  
\usepackage{algorithm}      
\usepackage{algpseudocode}  

\AtBeginDocument{%
  \providecommand\BibTeX{{%
    \normalfont B\kern-0.5em{\scshape i\kern-0.25em b}\kern-0.8em\TeX}}}

\begin{document}

\selectlanguage{english}
\title{py360tool: Um Framework para Manipulação de Vídeo 360$^\circ$ \\ com Ladrilhos}

\author{Henrique D. Garcia}
\email{henriquedgarcia@gmail.com}
\affiliation{%
  \institution{Departamento de Engenharia El\'etrica \\ Universidade de Brasília}
  \city{Brasília}
  \state{Distrito Federal}
  \country{Brasil}
}
\author{Marcelo M. Carvalho}
\email{mmcarvalho@txstate.edu}
\affiliation{%
  \institution{Ingram School of Engineering \\ Texas State University}
  \city{San Marcos}
  \state{Texas}
  \country{USA}
  }

\begin{abstract}

The streaming of 360$^\circ$ videos is one of the most bandwidth-demanding virtual reality (VR) applications, as the video must be encoded in ultra-high resolution to ensure an immersive experience. To optimize its transmission, current approaches partition the spherical video into tiles, which are encoded at different bitrates and selectively delivered, based on the viewing direction of the user (viewport). The complexity of this architecture, which involves viewport prediction, tile selection, bit rate adaptation, and handling of parallel streaming, requires new tools to evaluate quality of experience (QoE) and quality of service (QoS), especially due to its interactive nature and low reproducibility. This work introduces py360tools, a Python library to handle tile-based 360$^\circ$ video streaming. The library automates key client-side tasks, such as spherical projection reconstruction, viewport extraction, and tile selection, facilitating the playback and simulation of streaming sessions. Furthermore, py360tools offers a flexible architecture, enabling efficient analysis of different projections and tiling strategies.

\end{abstract}


\maketitle

\selectlanguage{brazil} 
\section{Introdução}

\begin{figure}[htb]
    \centering
  \includegraphics[width=\linewidth]{"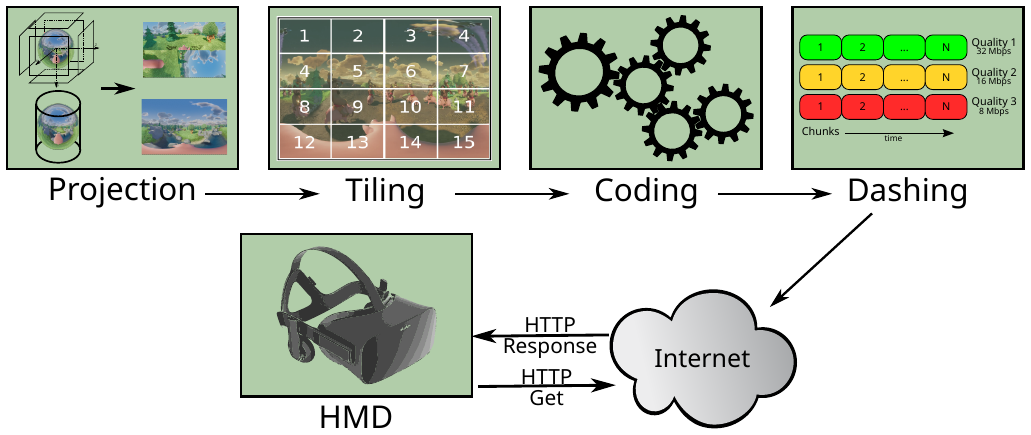"}
  \caption{Antes de ser transmitido o vídeo é projetado, recortado, comprimido e segmentado em \textit{chunks}. O cliente solicita apenas os \textit{chunks} que aparecem no \textit{viewport}}
  \label{fig:server-side}
\end{figure}

Dentre as aplicações de Realidade Virtual (RV), o \textit{streaming} de vídeo 360$^\circ$ é o que mais consome largura de banda. Como o campo de visão humano é limitado, é mais eficiente transmitir apenas a região da esfera que o usuário visualiza, chamada de \textit{viewport}. Uma das abordagens promissoras para otimizar essa transmissão é segmentar espacialmente a esfera em ladrilhos (\textit{tiles}) e transmitir apenas os que aparecem no \textit{viewport} usando protocolos como DASH (\textit{Dynamic Adaptive Streaming over HTTP}) ou HLS (\textit{HTTP live streaming}). O cliente então solicita apenas o \textit{chunk} contendo os ladrilhos visíveis naquele instante, adaptando-se à largura de banda disponível. A Figura~\ref{fig:server-side} ilustra o fluxo típico de processamento e transmissão de vídeo 360$^\circ$ via DASH: 
o vídeo 360$^\circ$ é projetado em um plano, segmentado em ladrilhos e codificado em diferentes qualidades/taxas de bits usando codificadores tradicionais (ex: H.265 ou VP9). Ao chegar no cliente, os \textit{chunks} de diferentes qualidades são remontados em cada quadro da projeção e intercalados de forma que o usuário veja o vídeo sem interrupções

A organização dos ladrilhos na projeção pode ser feita usando uma extensão do DASH chamada SRD (\textit{Spatial Relationship Description})~\cite{Dambra2018} ou outros métodos de identificação espacial dos ladrilhos que permite o envio de mais informações como \textit{JSON}~\cite{2025.araujo}. Porém, em todos os casos, o cliente deve saber onde colocar o ladrilho na esfera.
O vídeo é reproduzido em visualizadores de realidade virtual (\textit{Head-Mounted Display} -- HMD), que se conectam à CDN pela internet usando conexões sem fio como \textit{Wi-Fi} em uma arquitetura cliente-servidor. O cliente se conecta à CDN via protocolo HTTPS (\textit{HyperText Transfer Protocol Secure}) e, consequentemente, TCP (\textit{Transmission Control Protocol}). O cliente é responsável por controlar a taxa de bits de forma dinâmica, de acordo com a largura de banda disponível.

\begin{figure}[htb]
    \centering
    \includegraphics[width=1\linewidth]{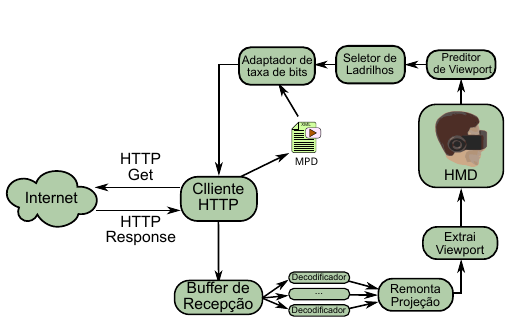}
    \caption{Internamente o cliente pode ser dividido em: preditor de \textit{viewport}, seletor de ladrilhos, ABR, Cliente HTTP, decodificador, montador de ladrilhos e extrator de \textit{viewport}. Estes módulos operam de forma sequencial e circular durante a reprodução do vídeo}
    \label{fig:client-side}
\end{figure}

O funcionamento de um cliente de streaming de vídeo 360$^\circ$ baseado em ladrilhos depende de múltiplos mecanismos, ilustrados na Figura~\ref{fig:client-side}.  
Inicialmente, via HTTP, o cliente requisita o manifesto que contém informações sobre os \textit{chunks}, localização espacial de cada ladrilho na projeção, taxa média de bits dos ladrilhos, respectivos URLs, e outras informações. Os segmentos recebidos são agendados para serem decodificados um pouco antes da sua reprodução, pois o vídeo sem compressão ocupa muita memória. Devido à codificação preditiva, todos os quadros de um {\em chunk} precisam ser totalmente decodificados para acessar um quadro arbitrário~\cite{Liu2019}. Antes de serem exibidos ao usuário, os ladrilhos decodificados são montados de volta na projeção e o \textit{viewport} é extraído.

Antes de solicitar o próximo {\em chunk}, o viewport é predito usando técnicas como regressão linear~\cite{Qian2016} ou aprendizado de máquina~\cite{Shao2025} e um mecanismo seletor de ladrilhos prioriza-os de acordo com as configurações do sistema~\cite{Nguyen2020}. É possível requisitar apenas os ladrilhos a serem vistos, ou requisitar ladrilhos com diferentes qualidades baseado na probabilidade de serem visualizados pelo usuário.  Por fim, o algoritmo de Adaptação da Taxa de Bits (ABR) define qual taxa de bits deve ser requerida ao servidor com base na prioridade, largura de banda prevista e informações do manifesto, buscando otimizar o uso de recursos e a experiência do usuário.

A natureza interativa do vídeo 360$^\circ$ dificulta a avaliação da qualidade subjetiva, pois a cada reprodução, os requisitos de transmissão dependem da direção para onde o usuário olha, não sendo possível submeter os participantes ao mesmo estímulo. Porém, podemos gravar o movimento de cabeça dos usuários assistindo o vídeo para então simular diferentes situações de transmissão considerando a mesma sequencia de viewports. Assim, esta nova sequência de viewport pode ser comparada com a {\em viewport} de referência ou ter sua qualidade avaliada de forma subjetiva.

Este trabalho apresenta a biblioteca {\tt py360tools}, desenvolvida em Python, para manipular vídeos 360$^\circ$ e seus elementos, como projeções, ladrilhos, e \textit{viewport}. A {\tt py360tools} encontra-se disponível no GitHub\footnote{\url{https://github.com/henriquedgarcia/py360tools}}. Esta ferramenta simula diversas etapas de um cliente de vídeo 360$^\circ$ com ladrilhos, servindo como base para um simulador de streaming de vídeo que permite avaliar a qualidade de experiência (\textit{Quality of Service} -- QoE) e qualidade de serviço (\textit{Quality of Service} -- QoS) sob diferentes condições de hardware e de rede.

\section{A biblioteca py360tools}

A py360tools é um conjunto de ferramentas sob licença GPL-3.0, projetadas para a pesquisa e simulação de streaming de vídeo 360$^\circ$ com ladrilhamento. A biblioteca oferece um conjunto de classes e funções para a manipulação geométrica das imagens, cujas principais funcionalidades são:

\begin{itemize}
    \item Seleção de Ladrilhos: Identificação e seleção dos ladrilhos que intersectam a \textit{viewport} do usuário em um dado instante;
    \item Reconstrução de Projeção: Montagem de uma projeção 2D completa a partir de um conjunto de ladrilhos individuais, que podem possuir diferentes qualidades;
    \item Extração de \textit{Viewport}: Geração da janela de visualização do usuário a partir de uma projeção 2D reconstruída;
    \item Conversão entre Projeções: Mapeamento de pixels entre diferentes formatos de projeção (ex: conversão entre projeções equirretangular e cúbica);
    \item Conversão de Ladrilhamento: Reorganização de um vídeo em diferentes esquemas de ladrilhamento (ex: alterando o número ou a disposição dos ladrilhos);
\end{itemize}

\begin{figure}[htp]
    \centering
    \includegraphics[width=1\linewidth]{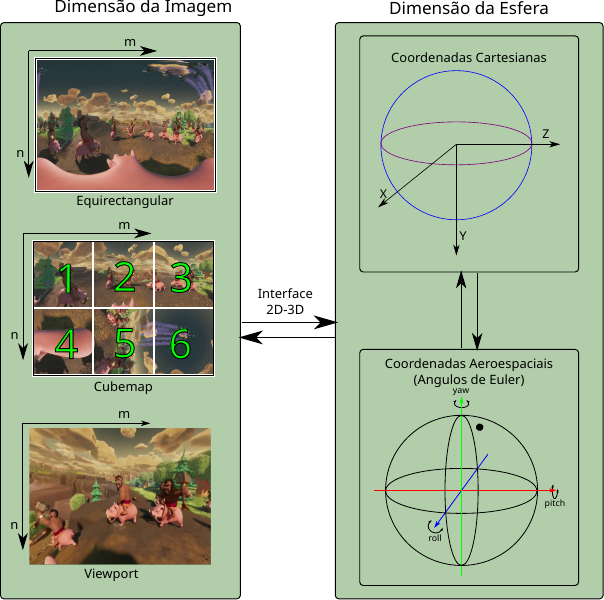}
     \caption{As transformações ocorrem em duas dimensões. A dimensão da esfera é manipulada em coordenadas cartesianas ou em ângulos de Euler. A dimensão da imagem é onde a codificação é efetuada e as distorções são geradas.}
    \label{fig:dimensoes}
\end{figure}

A Figura~\ref{fig:dimensoes} apresenta uma visão geral da arquitetura da biblioteca. O princípio básico da {\tt py360tools} é uma interface padrão para a conversão de coordenadas entre a dimensão da imagem (projeção 2D) e a dimensão da esfera (3D). Internamente, a biblioteca decodifica os ladrilhos recebidos e os utiliza para reconstruir a projeção. Em seguida, para qualquer operação de mapeamento, as coordenadas dos pixels são primeiro convertidas da projeção de origem para a esfera e, então, da esfera para a projeção de destino. Essa abstração permite o mapeamento flexível de pixels entre diferentes formatos de projeção e facilita operações complexas, como a extração precisa da \textit{viewport}.

A dimensão da imagem corresponde a um plano cartesiano 2D onde as coordenadas de cada pixel são representadas por valores inteiros em uma matriz. Nesta representação, todo conteúdo visual, incluindo as projeções esféricas (e.g., equirretangular) e a própria \textit{viewport}, é inicialmente tratado como uma imagem bidimensional. A dimensão da esfera atua como o espaço canônico para todas as transformações geométricas. Ela é definida por uma esfera de raio unitário centrada na origem de um sistema de coordenadas euclidiano 3D (conforme eixos ilustrados na Figura~\ref{fig:dimensoes}). Ao realizar uma conversão, os pixels da dimensão da imagem de origem são mapeados para vetores unitários neste espaço 3D. Estes vetores podem então ser remapeados de volta para a dimensão da imagem em um novo formato de projeção. A própria \textit{viewport} é tratada como um caso particular de projeção.

\subsection{O Modelo de Objeto}

\begin{figure}[htp]
    \centering
    \includegraphics[width=1\linewidth]{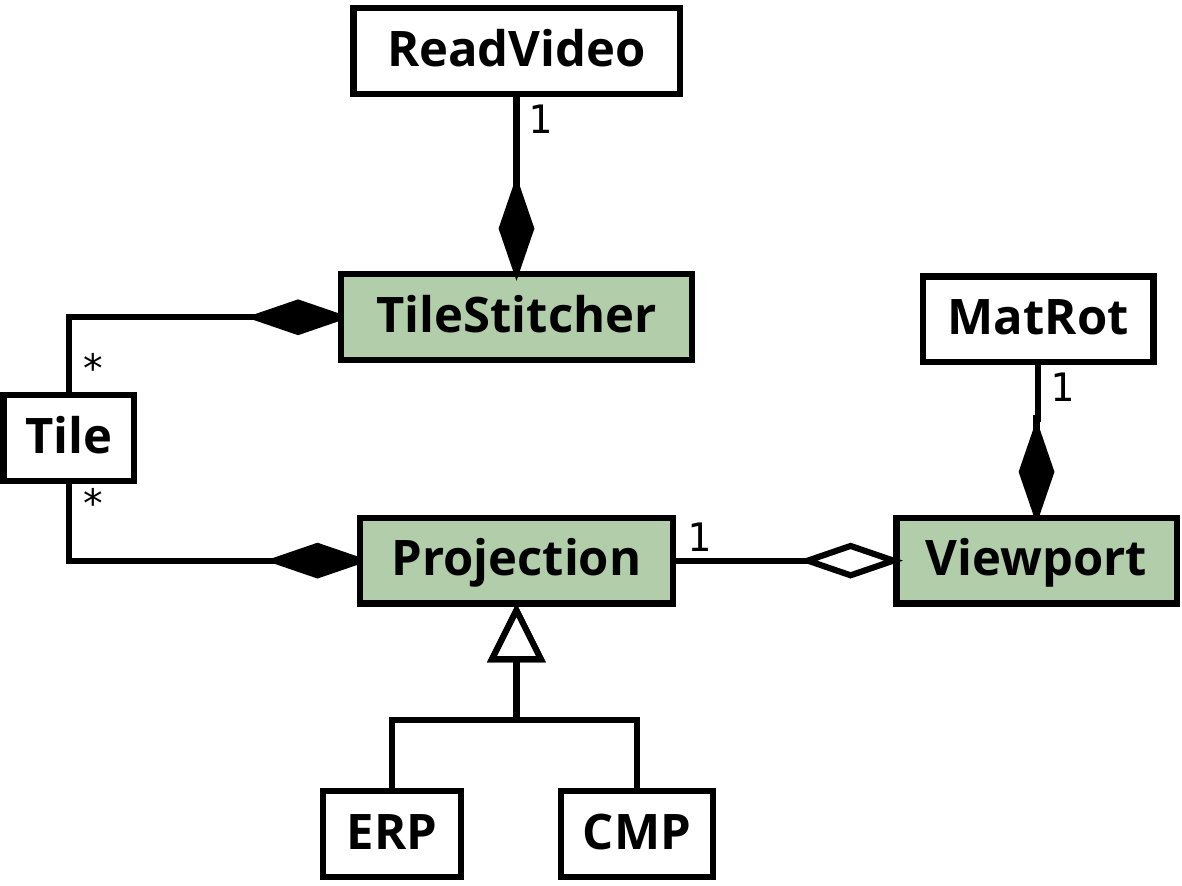}
    \caption{Diagrama de classes simplificado da \textit{py360tools}.}
    \label{fig:diagrama_de_classes}
\end{figure}

Para implementar estes conceitos foram criadas três classes principais (\textit{TileStitcher}, \textit{Projection} e \textit{Viewport}) cujo relacionamento é apresentado no diagrama simplificado da Figura~\ref{fig:diagrama_de_classes}.

A classe \textit{Projection} é uma classe abstrata para os diferentes formatos de projeção. Ela define operações na dimensão da esfera (3D), enquanto suas classes derivadas são responsáveis por implementar os mapeamentos entre a dimensão da projeção (2D) e da esfera (3D). Atualmente, a {\tt py360tools} oferece implementações para as projeções \textit{Cubemap} e \textit{Equirretangular}. Cada instância de \textit{Projection} possui uma lista de objetos \textit{Tiles}. A classe \textit{Tile} é uma \textit{dataclass} que armazena um índice, posição na projeção, tamanho e o caminho até o arquivo de vídeo associado ao ladrilho. 

A classe \textit{TileStitcher} é responsável por reconstruir a projeção 2D a partir de um dicionário de ladrilhos. Ela cria um objeto iterável que sincroniza e costura os ladrilhos fornecidos. Este objeto é inicializado com uma lista de \textit{Tiles}, o ladrilhamento e a resolução da projeção. Os vídeos dos ladrilhos são controlados por diversos objetos  \textit{ReadVideo}. Os objetos \textit{ReadVideo} são iteradores que iteram sobre os quadros de um vídeo que implementam funções para acessar o vídeo em uma posição especifica e reiniciar o vídeo.

De forma dinâmica, a cada iteração, o objeto \textit{TileStitcher} cria uma imagem vazia da mesma resolução que a projeção. Em seguida, para cada tile da lista, extrai um quadro e coloca em suas respectivas posições na projeção. Caso um ladrilho não esteja disponível, a área correspondente na projeção é preenchida com zeros e, por fim, retorna esta imagem.

Objetos da classe \textit{Viewport} fazem Agregação com \textit{Projection} e possuem duas funções principais: extrair o \textit{viewport} de uma projeção e determinar os ladrilhos visíveis. Para extrair o \textit{viewport}, esta classe implementa uma versão especializada da projeção Gnomônica, parametrizada pelo campo de visão (FoV) do HMD. A orientação da \textit{viewport} é atualizada dinamicamente para refletir a rotação da cabeça do usuário. Este processo é implementado criando a superfície do \textit{viewport} no espaço 3D e aplicando a matriz de rotação nos \textit{pixels} do \textit{viewport} baseada na direção que usuário está olhando. Em seguida convertemos estas novas coordenadas para o domínio da projeção e usamos interpolação para determinar a intensidade do pixel do viewport. 

A rotação da superfície do \textit{viewport} é definida pelo encadeamento de três rotações em torno dos eixos principais, utilizando os ângulos de Tait-Bryan, uma convenção particular dos ângulos de Euler. Estes ângulos são comumente conhecidos no contexto aeroespacial como (\textit{yaw}), (\textit{pitch}) e (\textit{roll}). A {\tt py360tools} adota como convenção de sinal as rotações no sentido horário sendo ângulos positivos. Esta convenção implica que as rotações dos ângulos de Tait-Bryan deve seguir a sequência Y-X-Z.

\begin{figure}[htb]
	\centering
	\includegraphics[width=0.9\linewidth]{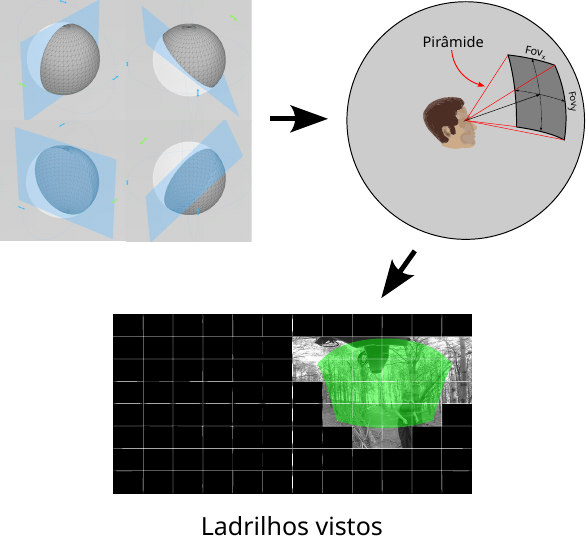}
	\caption{O que é visível (FOV) compreende a região do espaço que faz intersecção com quatro planos concêntricos. Basta um único pixel de um ladrilho estar dentro do FOV para que seja considerado visto.}
	\label{fig:makingviewport}
\end{figure}

Para determinar quais objetos no espaço 3D são visíveis, o campo de visão (\textit{Field of View} - FOV) do HMD é modelado como uma pirâmide de base quadrada cujo o ângulo de abertura determina a base. Esta pirâmide é geometricamente definida pela interseção de quatro planos que se encontram na origem da esfera (Figura~\ref{fig:makingviewport}). Os planos laterais (direito e esquerdo) formam o ângulo horizontal $FOV_x$, enquanto os planos superior e inferior definem o ângulo vertical $FOV_y$. Cada plano é representado por seu vetor normal, orientado para o interior da pirâmide. Consequentemente, a rotação do pirâmide é calculada aplicando-se a mesma matriz de rotação aos quatro vetores normais.

Um ponto \( p = \left( x, y, z  \right)  \) no espaço 3D pertence à \textit{viewport} se, e somente se, estiver no lado interno de todos os quatro planos. Matematicamente, esta condição é satisfeita se o produto escalar entre o vetor do ponto $ \vec{p} $ e o vetor normal $N_i=\left(x_i, y_i, z_i\right)$ de cada plano for negativo: $N_i \cdot \vec{p} \leq 0 $.
Para otimizar o processo de seleção de ladrilhos, não é necessário testar cada pixel individualmente. Em vez disso, o teste de pertencimento ao FOV pode ser aplicado de forma eficiente apenas às suas bordas.

Geralmente os dispositivos de realidade virtual, como o Meta Quest 3, possuem um FOV horizontal de aproximadamente 110$^\circ$ e um FOV vertical de 96$^\circ$~\footnote{https://www.meta.com/quest/quest-3/}. A resolução do \textit{viewport} depende do dispositivo. Quanto maior a resolução do \textit{viewport}, maior deve ser a resolução da projeção. Atualmente, os dispositivos apresentam visores com resolução de 2K (1080p) e suporte limitado a projeções em Ultra Alta Definição (UHD).

\begin{figure}[htb]
	\centering
	\includegraphics[width=1\linewidth]{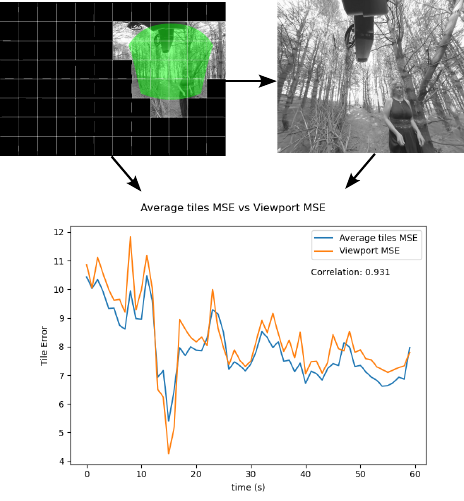}
	\caption{Não é necessário decodificar todos os ladrilhos para extrair o \textit{viewport}. Em vez disso, montamos a projeção apenas com os ladrilhos disponíveis e extraímos seu \textit{viewport}. Assim, podemos avaliar o quanto da distorção produzida pelo codificador ao ladrilho produz distorção no \textit{viewport} exibido ao usuário.}
	\label{fig:viewport}
\end{figure}

Como exemplo de aplicação, a Figura~\ref{fig:viewport} mostra o resultado da extração do \textit{viewport} da projeção preenchida apenas com alguns ladrilhos ao longo de uma sessão. Com base neste \textit{viewport} podemos avaliar a qualidade do que foi transmitido e quais recursos foram utilizados. Ao lado do \textit{viewport} temos um gráfico comparando o MSE do \textit{viewport} de um usuário com o MSE médio dos ladrilhos usados na sua reconstrução. A projeção foi segmentada no formato $12 \times8 $ e codificada com QP 28 e o \textit{viewport} foi extraído com FOV igual a $ 120^\circ \times 90^\circ$, com resolução de 1080p. 
Como esperado, o erro introduzido na codificação não produz o mesmo comportamento na projeção.

\section{Conclusões}

Este trabalho apresentou a {\tt py360tools}, uma biblioteca em Python de código aberto para a manipulação e simulação de \textit{streaming} de vídeo 360$^\circ$ baseado em ladrilhos. A ferramenta foi projetada para lidar com a complexidade geométrica de vídeos esféricos, oferecendo funcionalidades essenciais para a pesquisa em Realidade Virtual. Suas principais capacidades incluem a manipulação de ladrilhos, reconstrução de projeções, extração de \textit{viewports} e a conversão entre diferentes formatos de projeção. A arquitetura da {\tt py360tools} abstrai as operações geométricas entre as dimensões 2D (imagem) e 3D (esfera). Isso permite extrair e analisar métricas como taxa de bits, atraso de decodificação e qualidade do vídeo (QoE e QoS) com base nas estratégias de codificação e ladrilhamento.
Atualmente, a {\tt py360tools} é capaz de simular um ambiente ideal, com preditor de \textit{viewport} perfeito, ABR básico (qualidade constante), largura de banda infinita e atraso zero. Como trabalho futuro, pretendemos implementar um módulo de ABR e outro para predição de \textit{viewport}. Dessa forma, será possível integrar seu uso a simuladores de rede, como o NS-3, permitindo a simulação do sistema em um ambiente mais realista.
\bibliographystyle{ACM-Reference-Format}
\bibliography{sample-base}

\end{document}